\documentstyle[aps,preprint]{revtex}
\begin{document}
\draft


\preprint{TIT/HEP-422, YITP-99-33, gr-qc/9906047}

\title{Does the generalized second law require entropy bounds for a
charged system?}
\author{Takeshi Shimomura${}^{\dagger}$ and 
Shinji Mukohyama${}^{\ddagger}$}
\address{
${}^{\dagger}$Department of Physics, Tokyo Institute of Technology\\
Oh-Okayama, Meguro-ku,\\
Tokyo 152-0033, Japan \\
${}^{\ddagger}$Yukawa Institute for Theoretical Physics, 
Kyoto University \\
Kyoto 606-8502, Japan \\
${}^{\ddagger}$Department of Physics and Astronomy, 
University of Victoria \\
Victoria, BC, Canada V8W 3P6
}
\date{\today}

\maketitle


\begin{abstract} 

We calculate the net change in generalized entropy occurring when one
carries out the gedanken experiment in which a box initially
containing energy $E$, entropy $S$ and charge $Q$ is lowered
adiabatically toward a Reissner-Nordstr\"{o}m black hole and then
dropped in. This is an extension of the work of Unruh-Wald to a
charged system (the contents of the box possesses a charge $Q$). 
Their previous analysis showed that the effects of acceleration
radiation prevent violation of the generalized second law of
thermodynamics. In our more generic case, we show that the properties
of the thermal atmosphere are equally important when charge is
present. 
Indeed, we prove here that an equilibrium condition for the the
thermal atmosphere and the physical properties of ordinary matter are
sufficient to enforce the generalized second law. Thus, no additional
assumptions concerning entropy bounds on the contents of the box need
to be made in this process. 
The relation between our work and the recent works of Bekenstein and
Mayo~\cite{Bekenstein-Mayo}, and Hod~\cite{Hod} (entropy bound for a
charged system) are also discussed.

\end{abstract}
\pacs{PACS number(s): 04.70.Dy}

\section{Introduction}

One of the most remarkable developments in black hole physics 
is the relationship between the laws of black hole mechanics and  
thermodynamics.
Classically, black holes obey the laws that are 
analogous to the ordinary laws of thermodynamics~\cite{Bardeen}.
This correspondence becomes more than just an analogy
when quantum effects are taken into account
(Hawking's discovery of the thermal radiation emitted 
by a black hole~\cite{Hawking1975}).

Furthermore, Bekenstein~\cite{Bekenstein1973} has conjectured a 
generalized second law (GSL) of thermodynamics:
The sum of the black hole entropy and 
the ordinary entropy of the matter outside the black hole 
never decreases. 
More precisely, the GSL states that the generalized entropy 
$S_{g}$ defined by
%
\begin{equation}
  S_{g}  =  S_{matter} + {1\over 4} A_{bh}
\end{equation}
never decreases for any physical process
(we use natural units such that $\hbar=G=c=k=1$
throughout this paper),
where $S_{matter}$ is the entropy of ordinary matter outside 
the black hole and $A_{bh}/4$, one quarter of the
surface area of the black hole, plays the role of the entropy
of the black hole.
It is important to check the validity of this conjecture 
because this would strongly support the idea that 
the ordinary laws of thermodynamics apply to a self-gravitating 
quantum system containing a black hole and that $A_{bh}/4$ truly 
represents the physical entropy of the black hole.

There currently exists no general proof of the GSL based on the known 
microscopic laws of physics, although there are some proofs that
rely on the semiclassical
approximation~\cite{Frolov&Page,Sorkin,Mukohyama}. 
This is because the laws of quantum gravity are not well known. 
Thus, gedanken experiments to test the validity of the GSL are very
important tools to bolster confidence in this conjecture.

Classically, It was already recognized that a promising possibility 
for achieving a violation of the GSL occurs when one slowly (adiabatically) 
lowers a box initially containing energy $E$ and entropy $S$ toward a 
black hole and then dropped in~\cite{Bekenstein1973}.
The energy delivered to the black hole can be 
arbitrarily red-shifted by letting the dumping point approach
the horizon. Near this limit, the black hole area increase is not
large enough to compensate for the decrease of the matter's entropy.
A resolution of this difficulty was proposed by Bekenstein, who
conjectured that there exists a universal upper bound on the entropy
$S$ of matter with energy $E$ which is placed in a box of size
$R$~\cite{Bekenstein1981}:
%
\begin{equation}
 S  \le  2\pi E R. \label{eqn:bound}
\end{equation}
The intuitive reason why such a bound could rescue the GSL is that 
it prevents one from lowering a box close enough to a black hole 
to violate the GSL.

However, Unruh and Wald~\cite{Unruh-Wald1982,Unruh-Wald1983}
pointed out that Bekenstein failed to take into account certain
quantum effects in his analysis.
They noted that there is a quantum thermal atmosphere
surrounding a black hole, which produces a buoyancy force on a box
when one tries to lower the box slowly toward the black hole.
As a result, one cannot lower the box down to the horizon
(if one does not wish to inject energy by pushing it in)
and the box will float at a finite distance from the horizon,
which is determined by the condition that the energy contained 
in the box is exactly the same as the energy of the acceleration 
radiation displaced by the box.
Since the total energy at infinity added to the black hole after the 
box has been dropped from the floating point is larger than the 
redshifted proper energy of the box,
the box must be opened (this was extended to the {\lq}{\lq}dropped"
case, recently~\cite{Pelath-Wald}) at the floating point in order to 
minimize the entropy increase of the black hole.
Accordingly, they concluded that the GSL holds in this process provided 
only that unconstrained thermal matter maximizes entropy at fixed
volume and energy: 
%
\begin{equation}
 S  \le  V s(e),
\end{equation}
where $s(e)$ is the entropy density as a function of energy density $e$
of unconstrained thermal matter.
Thus, they concluded that no additional assumption on the quantum nature 
of the matter such as~(\ref{eqn:bound}) is necessary to rescue the GSL.

Recently, Bekenstein and Mayo~\cite{Bekenstein-Mayo} and 
Hod~\cite{Hod} have derived an upper bound to the entropy 
of a charged system by considering the polarization of the black hole
by a nearby charge.
They argued that the GSL could be saved only by assuming the existence
of entropy bounds on confined systems of the type as stated above.
In their derivation, they regard the system as a {\lq}{\lq}point particle"
and used the test particle approximation.
That is, the system is assumed to follow the equation of motion of 
a charged particle on a black hole background
and has a conserved energy
(the {\lq}{\lq}backreaction" effects are negligible).
However, since the system does not descend slowly (adiabatically) to
the black hole in this process, there must be backreaction effects:
the system radiates gravitational and electromagnetic radiation 
(these process also carry entropy) and the generalized entropy should
increase if all these effects are included. 
Further more, there is no justification for treating the system as a
point particle: 
the thermodynamical properties in and outside the box are 
completely neglected, even though they play an important role in the
validity of the GSL~\cite{Unruh-Wald1982,Unruh-Wald1983,Pelath-Wald}.
Thus, it is doubtful if this composite system can be considered 
to be thermal.

In order to avoid these difficulties,
we carry out a gedanken experiment in which a (possibly {\lq}{\lq}thick") 
box initially containing energy $E$, entropy $S$ and charge $Q$ is lowered 
adiabatically toward a Reissner-Nordstr\"{o}m black hole and 
then dropped in.
This is an extension of the work of Unruh-Wald 
to a charged system
(the contents of the box possess a charge $Q$).
Their previous analysis showed that the effects of acceleration 
radiation (buoyancy force) prevent a violation of the GSL, as stated above. 
Here, in addition to adding charge to the box, we consider the more
generic case in which the thermal atmosphere has a spherically
distributed charge, too. 
In this case, we notice that, in addition to the Unruh-Wald 
entropy restriction, there is an equilibrium condition for the
chemical potential of the thermal atmosphere. 
Indeed, we prove here that these two equilibrium conditions 
and the physical properties of ordinary matter are sufficient 
to enforce the generalized second law. Thus, no additional assumptions
concerning entropy bounds on the contents of the box need to be made
in this process.

In Sec.II, we derive the equations that hold for the 
thermal atmosphere around a black hole.
In Sec.III, we show that the GSL holds in the aforementioned process. 
Sec.IV is devoted to a summary and discussion of our results and, in
particular, comparison with previous
works~\cite{Bekenstein-Mayo,Hod}. 

\section{Thermal atmosphere around a black hole}

We carry out a gedanken experiment with a Reissner-Nordstr\"{o}m black
hole of mass $M$ and charge $Q_{bh}$, whose spacetime metric and
electromagnetic vector potential are given by
%
\begin{eqnarray}
 ds^2 & = & - f(r) dt^2 + \frac{dr^2}{f(r)} + r^2d\Omega^2,\\
 A_{\mu} dx^{\mu} & = & \Phi(r) dt \equiv -{Q_{bh}\over r} dt,
 \label{eqn:elemagpot}
\end{eqnarray}
where 
%
\begin{equation}
 f(r)  =  { (r-r_+) (r-r_-) \over r^2} ,
\end{equation}
with $r_{\pm}\equiv M\pm\sqrt{M^2-Q^2_{bh}}$.
The event horizon is located at $r=r_+$ and 
has area $A=4\pi {r^2_+}$.

The temperature of the black hole is defined by~\cite{Hawking1975}
%
\begin{equation}
 T_{H} = \frac{1}{2\pi}\kappa \equiv\frac{1}{4\pi}f^{\prime}(r_+),
\end{equation}
where $f^{\prime}$ denotes $df/dr$ and $\kappa
=(r_+ - r_-)/2{r^2_+}$ is the surface gravity of the
black hole.
Physically this represents the temperature of the black hole measured
at infinity.

First, we give a definition of unconstrained thermal matter with
charge.
We define unconstrained thermal matter in a given region 
outside the black hole to be the state of matter 
that maximizes entropy at a fixed volume, energy and charge
(electromagnetic potential given by~(\ref{eqn:elemagpot})).
Note that the properties of unconstrained thermal matter 
depend on location, i.e., the entropy density of unconstrained 
thermal matter, $\tilde{s}$, is a function of energy density, 
$\tilde{\rho}$,  charge density, $\tilde{q}$,
at the given point outside the black hole.
We assume that the thermal atmosphere of a black hole is described by
unconstrained thermal matter.

Then, the local temperature of a thermal atmosphere which is in
equilibrium with the black hole is given by using the Tolman's
law~\cite{Tolman} as 
%
\begin{equation}
 \tilde{T} = T_{H}/ \chi,	\label{eqn:Tolman-law}
\end{equation}
where $\chi=f^{1/2}$ is the redshift factor.

In addition to~(\ref{eqn:Tolman-law}), the 
chemical potential of the thermal atmosphere $\tilde{\mu}_i$
must satisfy the following condition
in order that it be in an equilibrium state~\cite{Landau-Lifshitz}:
%
\begin{equation}
 \tilde{\mu}_i \chi = {\mbox {Constant~for~each}}~i,	\label{eqn:chemical}
\end{equation}
where index $i$ denotes particle species.

Following Unruh and Wald~\cite{Unruh-Wald1982,Unruh-Wald1983},
we assume that the black hole has reached thermal equilibrium with 
the radiation, the whole system being enclosed in a large cavity.
This is achieved by fixing the boundary condition at the boundary,
i.e., by specifying the temperature and the electrostatic potential
(these are determined by the Hawking radiation and the difference 
between the chemical potentials, respectively) at the boundary.

The first law for the thermal atmosphere is written as 
%
\begin{equation}
 d\tilde{\rho} = \tilde{T}d\tilde{s} + \tilde{q}d\phi+
 \sum_{i}\tilde{\mu}_{i}d\tilde{n}_i,
\end{equation}
where $\phi=\Phi/\chi$. The integrated Gibbs-Duhem 
relation~\cite{Reif} for this system is as follows. 
(See appendix~\ref{app:Gibbs-Duhem} for a derivation.)
%
\begin{equation}
 \tilde{\rho} = \tilde{T}\tilde{s} - \tilde{P}+\sum_i\tilde{\mu}_i\tilde{n}_i.
	\label{eqn:Gibbs-Duhem}
\end{equation}
Note that the quantity $\phi$ does not appear in this expression.

By using the above two equations, the following equation is derived 
from Eqs.~(\ref{eqn:Tolman-law}) and (\ref{eqn:chemical}).
%
\begin{equation}
 d(\tilde{P}\chi) = -\tilde{\rho}d\chi - \tilde{q}\chi d\phi.
	\label{eqn:dPxhi}
\end{equation}
Eq.~(\ref{eqn:dPxhi}) states that pressure gradient is balanced by
gravitational and electromagnetic forces. 

\section{Validity of the generalized second law}

In this section, following~\cite{Unruh-Wald1983,Pelath-Wald,Bekenstein1994},
we compute the change in generalized entropy occurring when matter
in a (possibly {\lq}{\lq}large") box is slowly lowered toward a black hole 
and then dropped in.
We consider a box of cross-sectional area $A$ and height $b$, which 
contain energy density $\rho$, charge density $q$ and total entropy $S$.
As the box is lowered toward the black hole, the energy and charge density 
will depend both on the height $l$ of the center of the box above the horizon,
and the position within the box, $y$, as measured from the center.

We adopt the following notation for integrals~\cite{Bekenstein1994}
%
\begin{equation}
 \int f(y)dV \equiv A \int_{-b/2}^{b/2}f(y)dy.
\end{equation}
The energy of the box as measured at infinity is
%
\begin{equation}
 E_{\infty}(l) = \int \rho(l,y)\chi(l+y)dV, \label{eqn:energy}
\end{equation}
whereas the gravitational and electromagnetic forces 
as measured at infinity are in the forms
%
\begin{eqnarray}
 w(l) & = & \int \rho(l,y) \frac{\partial\chi(l+y)}{\partial l}dV,\\
 f_{em}(l) & = & \int q(l,y)\chi(l+y)
	\frac{\partial\phi(l+y)}{\partial l}dV.
\end{eqnarray}

These external forces do work on the gas in the box. We denote the
work by $W_{ge}(l)$:
%
\begin{eqnarray}
 W_{ge}(l) & = & E_{\infty}(l) - E_i  \label{eqn:gework}\\
 & = & \int_{\infty}^{l}
	[w(l^{\prime})+f_{em}(l^{\prime})]dl^{\prime},
\end{eqnarray}
where $E_i$ is the initial energy of the box. This is equivalent to 
%
\begin{equation}
 dE_{\infty} = (w+f_{em})dl. \label{eqn:dEdl}
\end{equation}

Meanwhile, the buoyancy force acting on the box, 
as measured at infinity, is equal to 
%
\begin{equation}
 f_b(l)  =  A\left[ (\tilde{P}\chi)_{l-b/2}
	-(\tilde{P}\chi)_{l+b/2}\right],
	\label{eqn:forces}
\end{equation}
where $\tilde{P}$ is the radiation pressure of the thermal
atmosphere.
From Eq.~(\ref{eqn:forces}), it is easy to show
that the work done by the buoyancy force is given by
%
\begin{eqnarray}
 W_b(l) = \int_{\infty}^{l}f_b(l^{\prime})dl^{\prime} 
            =  \int \tilde{P}(l,y)\chi(l+y)dV.
 \label{eqn:bwork}
\end{eqnarray}

Putting together Eqs.~(\ref{eqn:gework}) and (\ref{eqn:bwork}),
the total work done on the box system is given by 
%
\begin{eqnarray}
 W_{tot}(l) &=& W_{ge}(l)+W_{b}(l)\\
 	  &=& \int [\rho(l,y)+\tilde{P}(l,y)]\chi(l+y)dV - E_i.
 	  \label{eq:totalwork}
\end{eqnarray}

If the contents of the box are dropped into the black hole from
position $l_0$, the first law of black hole requires that the change
$\Delta S_{bh}$ in black hole entropy should satisfy
%
\begin{eqnarray}
 \Delta S_{bh} &=& \frac{1}{T_H}(E_i + W_{tot}(l_0) - \Phi_{bh}Q) \\
 	       &=& \frac{1}{T_H}
	\int [\rho(l_0,y)+\tilde{P}(l_0,y)]\chi(l_0+y)dV 
	- \frac{\Phi_{bh}Q}{T_H} ,
	\label{eqn:TdS}
\end{eqnarray}
where $Q$ and $\Phi_{bh} = Q_{bh}/r_+$ are charge thrown into the system by
the agent at infinity and electromagnetic potential of the black hole, 
respectively. 

Hereafter, for simplicity, we consider $2$-component system to be 
composed of a gas of particles with charge $e$ and anti-particles 
with opposite charge $-e$. 
This assumption does not affect our result and it is easy to
extend our argument to the $2n$-component system
\footnote{
If there is a particle with charge $e(>0)$, there exist a corresponding 
anti-particle with opposite charge $-e$ in nature.
So we consider an even number of particle species.
}, if we wish.

Therefore, by substituting Eqs.~(\ref{eqn:Tolman-law}) and
(\ref{eqn:Gibbs-Duhem}) into Eq.~(\ref{eqn:TdS}), we get the change in
generalized entropy as
%
\begin{eqnarray}
\Delta S_{g} &=& \Delta S_{bh} -S \\
     	    &=& \frac{1}{T_H}
	\int [\rho(l_0,y)-\tilde{\rho}(l_0,y)]\chi(l_0+y)dV
	+ \tilde{S}(l_0) - S\nonumber \\ 
	&{}{}{}{}{}& + \frac{1}{T_H}\left\{
	\sum_{i={+,-}}
	\int \tilde{\mu}_i(l_0,y)\tilde{n}_i(l_0,y)\chi(l_0+y) dV
	- \Phi_{bh}Q\right\},
	\label{eqn:DS-3}
\end{eqnarray}
where $\tilde{S}(l_0)=\int\tilde{s}(l_0,y)dV$ and $S=\int s(l_0,y)dV$
are the entropy of the thermal atmosphere displaced by the box 
and the entropy of the matter in a box,
respectively.

By using the equilibrium condition for the chemical potential
of the thermal atmosphere~(\ref{eqn:chemical}), and
noting that the chemical potential in the absence of the field can 
be neglected completely at the horizon because of 
its high (infinite) local temperature~\cite{early universe}, 
we get
%
\begin{eqnarray}
\tilde{\mu}_{+}\tilde{n}_{+}\chi+\tilde{\mu}_{-}\tilde{n}_{-}\chi
	&=&\tilde{\mu}^{h}_{+}\chi_{h}\tilde{n}_{+}
	+\tilde{\mu}^{h}_{-}\chi_{h}\tilde{n}_{-} \nonumber \\
	&=&e\Phi_{bh}(\tilde{n}_{+}-\tilde{n}_{-}),
	\label{eqn:chemical-condition}
\end{eqnarray}
where the index $h$ denotes the quantity evaluated at the horizon.

Thus, we can rewrite Eq.(\ref{eqn:DS-3}) further.
%
\begin{eqnarray}
 \Delta S_{g} &=& \int \left\{\tilde{s}(l_0)-
 		{1 \over T_H}[\tilde{\rho}(l_0,y)\chi(l_0+y)-\Phi_{bh}
 		\tilde{q}(l_0,y)]\right\}dV\nonumber \\ 
	&{}{}{}{}{}&-\int \left\{{s}(l_0)-{1 \over T_H}
		[{\rho}(l_0,y)\chi(l_0+y)-\Phi_{bh}{q}(l_0,y)] \right\}dV,
	\label{eqn:DS-4}
\end{eqnarray}
where $q=e(n_{+}-n_{-})$ and $\tilde{q}=e(\tilde{n}_{+}-\tilde{n}_{-})$
are the charge density of the matter in the box and 
that of the thermal atmosphere, respectively.

Since Eq.~(\ref{eqn:dEdl}) can be rewritten as 
%
\begin{equation}
 \int \frac{\partial\rho(l,y)}{\partial l}\chi(l+y)dV 
	- \int q(l,y)\chi(l+y)\frac{\partial\phi(l+y)}{\partial l}dV 
	= 0,
\end{equation}
it is easily shown by differentiating (\ref{eqn:DS-3}) that
%
\begin{equation}
 \frac{\partial}{\partial l_0}\Delta S_{bh} = 
	\frac{1}{T_H}\int\left\{
	[\rho(l_0,y)-\tilde{\rho}(l_0,y)]
	\frac{\partial\chi(l_0+y)}{\partial l_0}
	+ [q(l_0,y)-\tilde{q}(l_0,y)]\chi(l_0+y)
	\frac{\partial\phi(l_0+y)}{\partial l_0}\right\}dV,
	\label{eqn:DEF}
\end{equation}
where we have used Eqs.~(\ref{eqn:chemical}) and (\ref{eqn:dPxhi}),
and the fact that total charge $Q=\int qdV$ of the particles in the
box is conserved. Note that, since this process of lowering the box is
adiabatic, no change in the entropy in the box can occur and thus
$\Delta S_{bh}$ can be replaced by $\Delta S_{g}$.

First, for simplicity, we consider the case in which the box is 
sufficiently {\lq}{\lq}small" in the sense that the change in $\chi$,
$d\chi/dl$ and $d\Phi/dl$ across the box are small compared with 
their average values.

In this case, the floating point condition (\ref{eqn:DEF}) and the
total change in generalized entropy (\ref{eqn:DS-4}) reduce to 
%
\begin{equation}
	\rho(l_0) + q(l_0)\chi(l_0)
	\frac{\partial\phi(l_0)}{\partial \chi(l_0)}
	=\tilde{\rho}(l_0) + \tilde{q}(l_0)\chi(l_0)
	\frac{\partial\phi(l_0)}{\partial \chi(l_0)},
	\label{eqn:CONST}
\end{equation}
and
%
\begin{equation}
 \Delta S_{g} = \left\{ \tilde{S}(l_0)-
 		\frac{1}{T_H}[\tilde{E}(l_0)-\tilde{Q}(l_0)\Phi_{bh}]
 		\right\}
 		-\left\{ {S}(l_0)-
 		\frac{1}{T_H}[{E}(l_0)-{Q}(l_0)\Phi_{bh}]
 		\right\},
	\label{eqn:DS-5}
\end{equation}
respectively. Here, we wrote $S=sV$, $E= \rho\chi V$, $Q=qV$ and the
quantities with $(~\tilde{}~)$ refer to the thermal atmosphere.

Thus, our task is to seek the distribution function which
maximizes the functional $S-(E-Q\Phi_{bh})/T_{H}$ under the
constraint~(\ref{eqn:CONST}).

The state of matter is encoded in some density operator $\hat{f}$.
By using it, we can express the energy density as 
$\rho=Tr(\hat{\rho}\hat{f})$  
and charge density as $q=Tr (\hat{q}\hat{f})$, while entropy is
defined by $sV= -Tr (\hat{f}\ln{\hat{f}})$. 
Then, Eq.~(\ref{eqn:CONST}) can be rewritten as
%
\begin{equation}
	Tr[\hat{O}\hat{f}] = Tr[\hat{O}\tilde{f}],
	\label{eqn:CONST2}
\end{equation}
where $\hat{O}\equiv\hat{\rho}\chi V + 
\hat{q}\chi^2{\partial\phi \over \partial\chi}V$, $\hat{f}$ and $\tilde{f}~
(\propto\exp{\{-(\tilde{H}_{\infty}-\tilde{Q}\Phi_{bh})/~T_{H}\}})$
denotes the density operator of the matter in a box and that of the 
thermal atmosphere which is in equilibrium with the black hole,
respectively.

Considering that the variation of $S-(E-Q\Phi_{bh})/T_{H}$
under a small variation $\delta\hat{f}$ is given by
%
\begin{eqnarray}
 \delta [S-(E-Q\Phi_{bh})/T_{H}] 
 	&=& -Tr[\delta{\hat{f}}
 	(\ln{\hat{f}}+1+(\hat{\rho}\chi-\hat{q}\Phi_{bh})V{T_{H}}^{-1})] 
 	\nonumber \\
 	&\equiv& -Tr[\delta{\hat{f}}
 	(\ln{\hat{f}}+1+(\hat{H}_{\infty}-\hat{Q}\Phi_{bh}){T_{H}}^{-1})],
	\label{eqn:variation}
\end{eqnarray}
the functional $S-(E-Q\Phi_{bh})/T_{H}$ has an extremum under 
variations that preserve
$Tr{\hat{f}}=1$ and $Tr[\hat{O}\hat{f}] = Tr[\hat{O}\tilde{f}]$, 
where $\hat{f}$ satisfies
$(\ln{\hat{f}}+1)+(\hat{H}_{\infty}-\hat{Q}\Phi_{bh}){T_{H}}^{-1}
-\lambda_{1}-\lambda_{2}\hat{O}=0$.
The quantities $\lambda_{1,2}$ are Lagrange multipliers 
for these constraints.
Eliminating $\lambda_{1}$ by using $Tr{\hat{f}}=1$, 
we get a unique solution
%
\begin{equation}
 \hat{f} = \frac{1}{Z}
 	\exp{\left\{-\beta_H(\hat{H}_{\infty}-\hat{Q}\Phi_{bh})
 	+\lambda_{2}\hat{O}\right\}},
	\label{eqn:dis-fn}
\end{equation}
where $Z=Tr[\exp{\{-\beta_H(\hat{H}_{\infty}-\hat{Q}\Phi_{bh})
+\lambda_{2}\hat{O}\}}]$ and $\beta_{H}\equiv T_{H}^{-1}$. 
Then, by substituting Eq.~(\ref{eqn:dis-fn}) into Eq.~(\ref{eqn:CONST2}), 
we get $\lambda_{2}=0$ and thus $\hat{f}=\tilde{f}=
Z^{-1}\exp{\{-\beta_H(\hat{H}_{\infty}-\hat{Q}\Phi_{bh})\}}$.

Therefore, the maximum value of the functional
$S-(E-Q\Phi_{bh})/T_{H}$  is realized for the thermal state 
with the canonical distribution 
$\hat{f}=Z^{-1}\exp{\{-\frac{1}{T_{H}}
(\hat{H}_{\infty}-\hat{Q}\Phi_{bh})\}}$,
which, in our case, corresponds to the thermal atmosphere 
of the black hole. 

Hence, we have
%
\begin{equation}
	\Delta S_{g} \ge \Delta S_{g}(l=l_0)\ge0.
\end{equation}
Thus, the GSL is satisfied in this process.

Next, we analyze the case of a {\lq}{\lq}larger" box.
The same procedure as for the {\lq}{\lq}small" box can be
applied in this case, too.

Hereafter, we adopt the following notation
%
\begin{equation}
	\int adV \equiv 
	Tr[\hat{A}\hat{f}],
	\label{eqn:CONST3}
\end{equation}
where $a$, $\hat{A}$ and $\hat{f}$ denote some observable,
corresponding operator and density operator, respectively.

With this notation, the total change in generalized entropy 
(\ref{eqn:DS-4}) can be written in the form
%
\begin{equation}
 \Delta S_{g} = U[\tilde{f};\beta_{H},\Phi_{bh}]
		-U[\hat{f};\beta_{H},\Phi_{bh}],
	\label{eqn:DS-L}
\end{equation}
where $U$ is a functional of a density matrix of the matter fields
defined by
%
\begin{equation}
 U[\hat{f};\beta_{H},\Phi_{bh}]
		\equiv
		- Tr[\hat{f}\ln{\hat{f}}]-{\beta_{H}}(Tr[\hat{H}_{\infty}
		\hat{f}]-\Phi_{bh}Tr[\hat{Q}\hat{f}]),
	\label{eqn:DS-def}
\end{equation}
$\hat{f}$ and $\tilde{f}~
(\propto\exp{\{-\beta_{H}(\tilde{H}_{\infty}-\tilde{Q}\Phi_{bh})\}})$
denote the density operators of the matter in the box and of the 
thermal atmosphere which is equilibrium with the black hole,
respectively. 
In this expression, $\hat{H}_{\infty}\equiv\int\hat{\rho}\chi dV$ and 
$\hat{Q}\equiv\int \hat{q} dV$ are operators corresponding to energy 
(at infinity) and charge.
Note that the functional $U$ is essentially the negative of the free
energy divided by the temperature.

Similarly, the floating point condition (\ref{eqn:DEF}) can be reduced 
to 
%
\begin{eqnarray}
  Tr[\hat{O}\tilde{f}]&=&Tr[\hat{O}\hat{f}] \label{eqn:CONST-L}\\
		&\equiv&\int(\rho\frac{\partial\chi}{\partial l}
	        +q\chi\frac{\partial\phi}{\partial l})dV.
\end{eqnarray}
These Eqs.~(\ref{eqn:DS-L}) and (\ref{eqn:CONST-L}) have just the same
form as the Eqs.~(\ref{eqn:DS-5}) and (\ref{eqn:CONST2}) for 
the {\lq}{\lq}small'' box.
Therefore, by repeating the same procedure as in the {\lq}{\lq}small"
box's case, we can show that violation of the GSL cannot be achieved 
in the case of a {\lq}{\lq}large" box.

In obtaining these results, we have ignored any entropy emitted by the 
black hole.
In fact, the entropy produced in spontaneous Unruh emission 
corresponding to the superradiant modes can be neglected by taking 
the black hole as a very massive one.

\section{Summary and discussion}

We examined the gedanken experiment of lowering
a box initially containing energy $E$, entropy $S$ and charge $Q$
toward a Reissner-Nordstr\"{o}m black hole and then dropped in
(an extension of the work of Unruh-Wald to the charged system).
We have shown that the properties of the thermal atmosphere plays 
an important role in this case just as in Unruh-Wald's case.
Specifically, we used an assumption that unconstrained thermal matter 
maximizes entropy as a function of energy density and charge density,
in addition to the Unruh-Wald buoyancy force.
Note that an equilibrium condition for the chemical potential of 
the thermal atmosphere also plays an important role in this case.
In deed, we proved here that these are sufficient 
for the enforcement of the GSL
and no additional assumptions concerning entropy bounds 
on the contents of the box need to be made in this process.

Finally, we comment briefly on the relation between our work and 
the recent work of Bekenstein and Mayo~\cite{Bekenstein-Mayo}, and 
Hod~\cite{Hod}.
They have derived an upper bound to the entropy of a charged system 
by considering the polarization of the black hole by a nearby charge
(gravitationally induced electrostatic self-force on a charged 
test particle~\cite{Smith}).
They concluded that the GSL could be saved only by assuming 
the existence of entropy bounds on a confined charged system.
On the other hand, in our derivation, we have neglected the electrostatic 
self-force till now. If we want to include the electrostatic self-energy 
in our analysis, we have only to replace 
$\Phi\rightarrow \Phi\pm eM/{2 r^2}$ in our analysis. 
The only effect is a change in Eq.~(\ref{eqn:chemical-condition}),
i.e., 
%
\begin{equation}
\tilde{\mu}_{+}\tilde{n}_{+}\chi+\tilde{\mu}_{-}\tilde{n}_{-}\chi
	=e\Phi_{bh}(\tilde{n}_{+}-\tilde{n}_{-})
	+\frac{e^2 M}{2r_{+}^2}(\tilde{n}_{+}+\tilde{n}_{-}).
	\label{eqn:self-force}
\end{equation}
This correction gives a positive contribution to the net change 
in generalized entropy Eq.~(\ref{eqn:DS-4}). 
Thus, in this gedanken experiment, the GSL would hold even if we include
those self-interaction forces.

There are several advantages to our analysis compared with 
theirs. They regarded the system as a {\lq}{\lq}point particle"
(test particle approximation)
and assumed that it follows the equation of motion of a charged particle 
on a black hole background and has a conserved energy
(the {\lq}{\lq}backreaction" effects are negligible).
However, the system does not descend slowly (adiabatically)
to the black hole in this process,
the system would radiate gravitational and electromagnetic radiation 
(these process also carry entropy) and
the generalized entropy should increase if all these effects are
included. Of course, such an analysis including the backreaction
effect would be too complicated to reach a definitive answer
analytically. Compared this, since we very slowly (adiabatically)
lowers the box toward the black hole (quasi-static process), these
effects can be neglected. Furthermore, there is no justification for
treating the system as a point particle: 
the thermodynamical properties in and outside the box is 
completely neglected~\footnote
{
Thus, it is doubtful if their composite system can be considered 
to be a thermal one (in the sense of thermally contacted system).
On the other hand, since we adiabatically lowers the box 
toward the black hole, this condition is naturally justified.
},
even though they play an important role in the validity 
of the GSL~\cite{Unruh-Wald1982,Unruh-Wald1983,Pelath-Wald}.
In deed, we take into account the energy change in the box and 
the effects of thermal atmosphere and showed that
these effects have an important role to prevent the violation of the GSL.

Of course, our analysis is not perfect:
for instance, we have neglected interactions between the constituents
of the radiation and the thermal atmosphere.
However, we could say that our analysis improves the previous
analyses, even if we have not resolved all the difficulties.

\vskip 1cm
\centerline{\bf Acknowledgments}

One of us (T.S.) would like to thank Dr. T. Okamura for
useful discussions, Professors A. Hosoya, H. Ishihara and 
T. Mishima for their continuing encouragement. 
The other (S.M.) would like to thank Professor H. Kodama for his
continuing encouragement and Professor W. Israel for his warmest
hospitality in University of Victoria and careful reading of the 
manuscript. 
This work was supported partially (S.M.) by the Grant-in-Aid for
Scientific Research Fund (No. 9809228).

\appendix

\section{Integrated Gibbs-Duhem relation}
	\label{app:Gibbs-Duhem}

Providing that the system is in equilibrium states, 
the first law of thermodynamics is
%
\begin{equation}
 d{\cal E} = \tilde{T}d{\cal S} - \tilde{P}d{\cal V} + {\cal Q}d\phi
 	   +\sum_i\tilde{\mu}_id{\cal N}_i
 	   \label{app:first law},
\end{equation}
where ${\cal E}$, $\tilde{T}$, ${\cal S}$, $\tilde{P}$, ${\cal V}$, ${\cal Q}$,
${\phi}$ ,$\tilde{\mu}_i$ and ${\cal N}_i$
are energy measured by a local static observer, local temperature,
entropy, pressure, volume, electromagnetic charge, electromagnetic potential,
chemical potential and particle number density, respectively.

Since 
%
\begin{equation}
 d({\cal E}-{\cal Q}\phi) = 
	\tilde{T}d{\cal S} - \tilde{P}d{\cal V} - \phi d{\cal Q}
	+\sum_i\tilde{\mu}_id{\cal N}_i,
\end{equation}
the quantity ${\cal E}-{\cal Q}\phi$ is a function of ${\cal S}$,
${\cal V}$, ${\cal Q}$ and ${\cal N}_i$: 
%
\begin{equation}
 {\cal E}-{\cal Q}\phi = {\cal F}({\cal S},{\cal V},{\cal Q},{\cal N}_i).
\end{equation}
Thus, since ${\cal F}$ is homogeneous function of degree 1 in 
these extensive parameters, we get 
%
\begin{equation}
 \alpha({\cal E}-{\cal Q}\phi) = 
	{\cal F}(\alpha{\cal S},\alpha{\cal V},\alpha{\cal Q},\alpha{\cal N}_i).
\end{equation}
By differentiating this equation with respect to $\alpha$ and setting
$\alpha=1$, we obtain 
%
\begin{equation}
 {\cal E}-{\cal Q}\phi = 
	\left(\frac{\partial{\cal F}}{\partial{\cal S}}
	\right)_{{\cal V},{\cal Q},{\cal N}_i}{\cal S}
	+ \left(\frac{\partial{\cal F}}{\partial{\cal V}}
	\right)_{{\cal S},{\cal Q},{\cal N}_i}{\cal V}
	+ \left(\frac{\partial{\cal F}}{\partial{\cal Q}}
	\right)_{{\cal S},{\cal V},{\cal N}_i}{\cal Q}
	+ \sum_i\left(\frac{\partial{\cal F}}{\partial{\cal N}_i}
	\right)_{{\cal S},{\cal V},{\cal Q}}{\cal N}_i.
\end{equation}
Therefore, we get the integrated Gibbs-Duhem relation:
%
\begin{equation}
 {\cal E} = \tilde{T}{\cal S} - \tilde{P}{\cal V} 
 + \sum_i\tilde{\mu}_i{\cal N}_i
 	\label{app:GD rel}.
\end{equation}
Eqs.(\ref{app:first law}) and (\ref{app:GD rel}) also can be 
rewritten as the relations between local quantities:

%
\begin{eqnarray}
 d\tilde{\rho} & = & \tilde{T}d\tilde{s} + \tilde{q}d\phi + 
 \sum_i\tilde{\mu}_id\tilde{n}_i,
 \nonumber\\
 \tilde{\rho} & = & \tilde{T}\tilde{s} - \tilde{P}+ 
 \sum_i\tilde{\mu}_i\tilde{n}_i.
\end{eqnarray}
where $\tilde{\rho}$ ($={\cal E}/{\cal V}$), $\tilde{s}$ 
($={\cal S}/{\cal V}$), $\tilde{q}$ ($={\cal Q}/{\cal V}$) and
$\tilde{n}_i$ ($={\cal N}_i/{\cal V}$) are energy density measured 
by a local static observer, entropy density, charge density and number density,
respectively.




\begin{thebibliography}{22}
\bibitem{Bardeen}
J. M. Bardeen, B. Carter and S. W. Hawking, 
Commun. Math. Phys. {\bf 31},161 (1973).
\bibitem{Hawking1975}
S. W. Hawking, Commun. Math. Phys. {\bf 43}, 199 (1975).
\bibitem{Bekenstein1973}
J. D. Bekenstein, Phys. Rev. D{\bf 7}, 2333 (1973).
\bibitem{Frolov&Page}
V. P. Frolov, D. N. Page, Phys. Rev. Lett. {\bf 71}, 3902 (1993).
\bibitem{Sorkin}
R. D. Sorkin, Phys. Rev. Lett. {\bf 56}, 1885 (1986).
\bibitem{Mukohyama}
S. Mukohyama, Phys. Rev. D{\bf 56}, 2912 (1997).
\bibitem{Bekenstein1981}
J. D. Bekenstein, Phys. Rev. D{\bf 23}, 287 (1981).
\bibitem{Unruh-Wald1982}
W. G. Unruh and R. M. Wald, Phys. Rev. D{\bf 25}, 942 (1982).
\bibitem{Unruh-Wald1983}
W. G. Unruh and R. M. Wald, Phys. Rev. D{\bf 27}, 2271 (1983).
\bibitem{Pelath-Wald}
M. A. Pelath and R. M. Wald, gr-qc/9901032.
\bibitem{Bekenstein-Mayo}
J. D. Bekenstein and A. E. Mayo, gr-qc/9903002.
\bibitem{Hod}
S. Hod, gr-qc/9903010; gr-qc/9903011.
\bibitem{Tolman}
R. C. Tolman, {\it Relativity, Thermodynamics and Cosmology}
(Clarendon, Oxford, 1934), p. 318.
\bibitem{Landau-Lifshitz}
For example, L. D. Landau and E. M. Lifshitz, {\it Statistical Physics}
(Pergamon Press, 1980), p. 76.
\bibitem{Reif}
For example, F. Reif, {\it Fundamentals of statistical and thermal
physics} (McGraw-Hill, 1965), p. 315. 
\bibitem{Bekenstein1994}
J. D. Bekenstein, Phys. Rev. D{\bf 49}, 1912 (1994).
\bibitem{Anderson}
W. G. Anderson, Phys. Rev. D{\bf 50}, 4786 (1994).
\bibitem{early universe}
For example, E. W. Kolb and M. S. Turner, {\it The Early Universe}
(Addison-Wesley, 1993), footnote in p. 66.
\bibitem{Smith}
D. Lohiya, J. Phys. A: Math. Gen.{\bf 15}, 1815 (1982).

\end{thebibliography}
\end{document}